# Use of social networks to motivate computer-engineering students to participate in self-assessment activities

Carlos GUERRERO[1], Antoni JAUME-I-CAPÓ[2]

[1] University of Balearic Islands, Crta. Valldemossa km 7.5, Palma, E-07122, Spain, carlos.guerrero@uib.es

[2] University of Balearic Islands, Crta. Valldemossa km 7.5, Palma, E-07122, Spain, antoni.jaume@uib.es

**Abstract:** Motivation is essential in the learning process of university students, and teachers should have a wide range of strategies to address this issue. The emergence of social technologies has had a considerable influence in e-learning systems, and a number of experts state that their use is a good method to motivate students and to increase their participation in activities. This study attempted to determine if social networks and social applications are just another tool or if they provide extra motivation for students to participate. The study compared the percentage of student participation in tasks of self-assessment. The experiments covered three traditional strategies of student motivation and another one in which social networks were used to introduce, explain and deliver the self-assessment tasks. The case with a higher participation was when students obtained a reward from the completion of the activity. Despite this result, the statistical analysis indicated that the use of social networks obtained similar results as a strategy of continuous and regular motivation speeches.

**Carlos Guerrero**[1] received his Ph.D. degree in Computer Engineering at the Balearic Islands University in 2012. He is an assistant professor of Computer Architecture and Technology at the Computer Science Department of the University of the Balearic Islands. His research interests include web performance, web engineering, web applications, data mining, intelligent systems, mobile systems, educational innovation and use of web 2.0 in learning processes. He has authored around 12 papers in different international conferences and journals. He has been member of the program committee of several international conferences.

**Antoni Jaume-i-Capó**[2] received his Ph.D. degree in Computer Engineering at the Balearic Islands University in 2009. He has been an assistant professor at the Mathematics and Computer Science Department of the Balearic Islands University since 2005. His main research interests include computational vision, vision-based interfaces, serious games, rehabilitation and motivation. He has authored 8 publications in international journals and more than 15 papers in international conferences.

## 1. Introduction

The use of social computing and social networks has extended to a wide range of fields and activities. Education in general, and at universities in particular, is also influenced by this technological evolution, as shown by the numerous new applications that integrate learning processes and social activities. However, some professors are a bit skeptical about the usefulness of this type of technology. The research we present in this article attempts to determine the influence of social networks on student motivation. The goal was to study if social networks and social applications are just a new type of tool professors can use or if they are, in themselves, a way to motivate students.

The study presents quantitative research comparing the significant differences among four different motivation strategies. Student responses to these strategies were compared based on their participation in course activities. Self-assessment was the activity of interest because it is notably useful but also difficult to convince students to engage in it.

The number of students who actually performed the goal activity (self-assessment task) was measured in four different motivation cases. The first strategy, which constituted the control group, consisted of explaining to the students only once at the beginning of the activity why they should complete the self-assessment tasks (initial strategy). The second one was to remind the students periodically to self-assess, insisting often on its advantages (regular strategy). A third strategy was to reward through their grades those students who complete self-assessment tasks (rewarded strategy). Finally, we used social networks to convince students to self-assess (social strategy).

Every experienced instructor knows that the best possible motivation for student participation is to award points towards his grade for participation. It is also clear that more students complete a task if they are reminded about it multiple times than if they are only told once. Despite these influences, the numerical differences between these three cases are not clear. The first goal of this experiment was to measure the numerical differences between these three traditional motivation strategies. Moreover, the experiment was designed to establish the relative student participation achieved due to the use of social networks compared to the other three strategies.

The study was conducted on computer engineering students. The results showed that there is a significant and important improvement in using the social and the regular strategies over the initial one. These two (social and regular) did not show a statistically significant difference. As expected, the rewarded strategy showed the highest student participation. The results were validated by studying the significant differences using the Kruskal-Wallis test.

## 2. Related Work

Research efforts into the use of social computing and tools in e-learning environments have increased considerably during recent years. This research is focused on the idea that the use of social tools is an extra motivation for the students and it increases the participation of students in course activities. Several researchers have studied the effects of social networks on students. Thus, we present here recent papers related to student engagement and social, collaborative and e-learning web systems.

The emergence of information technologies was an education and learning processes revolution, improving some of the traditional methods through the use of web systems, mobile applications and similar technologies. Currently, a new type of web technology has emerged, mainly via the addition of a social component to their data flows and interactions. E-learning platforms are gaining huge advantages from these technologies. For example, there are many experiences that involve the use of social networks in subject development. There is plenty of evidence that these new technologies increase the motivation and engagement of students in the subjects and in the learning process.

Ellison et al. [1] studied the benefits of the use of social networks in terms of increased student satisfaction and in terms of a higher level of formation and maintenance of social capital and networking, which means improvements in student learning. Mazerd et al. [2] studied how the use of Facebook increased and predict the student motivation, affective learning and classroom climate. Susnea et al. [3] studied the behavior of learners' networks and modelled them using a P2P protocol. The authors demonstrated the economic benefits of collaborative learning processes.

Cheung et al. studied the factors that influence student use of social networks [4]. They determined that social presence is one of the most important values to their use. The professors need to take advantage of these inclinations, using social sites and including these systems in their teaching. Simões et al. [5] analyzed the interest of the students toward Personal Learning Environment (PLE) systems. However, one of the open problems with the use of social networking sites is that the professors have some reservations [6]. Studies such as the one we present in this article were addressed to avoid this unpleasant feeling when using social computing in classrooms and lectures.

Thoms presented a statistical study about the relationship between the use of e-learning support tools and the perceived levels of learning in the students [7]. This system allowed students to provide instant feedback to their peers in an on-line learning community. The study indicated a significant difference between groups with different tool usage levels.

Michalco et al. studied how to increase the participation of people in general activities using social networks [8]. The researchers performed quantitative and qualitative analyses. The main difference with our studied is that ours is focused on the learning environment and students instead of on a general environment.

In addition to these two last studies, it is quite usual to find experiences and studies about how useful social computing is in learning that lack a serious statistical study of the differences between the student groups. Our objective is to determine if the use of social networks, e-learning environments and similar technologies increase student participation in activities.

We limited the study to motivating students to complete self-assessment tasks. Assessment is a part of the learning process [9], [10]. Students acquire concepts in a deeper fashion and improve their learning management [11]. The feedback is very important for students, but high levels of feedback impose a high workload on teachers. There are some proposals to help teachers in the assessment process [12], [13], but they are not sufficient to balance the increase in the workload due to innovative learning methodologies where the feedback for the students is very important. Self-assessment is a good solution for this problem. However, the increase in the student feedback results in a high workload for the professors. Boud and Harvey proved that the use of peer and self-assessment was a good method to reduce the teacher workload in large classes and, at the same time, these methods provide educational benefits [14]. The use of e-learning systems can help reduce the workload further. For this reason, processes of self-assessment are usually related to the use of e-learning systems and tools.

Sondergaard analyzed the use of an on-line tool for student peer assessment and the response of the students to its use [15]. Other examples are Kybartaite et al. [16] and Gehringer [17]. The first one studies the benefits of virtual campus systems and the second one analyzed web systems for the peer review of students. Finally, Thorsteinsson et al. performed a qualitative study of ICT in education and Managed Learning Environments (MLE) [18]. The authors detected the teacher's problems when they use this type of environments.

More advanced works presented on-line tools that provide more detailed, informed and less biased assessment for the students peer's work to coordinate the cognitive differences between students [19]. The authors of this work, Lan et al., evaluated the students' willingness to accept the assessment results, but they did not compare the students' participation with other methods of assessment.

## 3. Method

The objective of the research was to study if the use of social computing and tools increased the student participation in activities. The study was limited to the motivation of the students in self-assessment tasks. Self-assessment was included in several student groups and different motivation strategies were applied to motivate students to complete these tasks. The percentage of students that took part in the activities was measured.

Therefore, the main hypothesis of the research is that student participation is not only related to the motivation strategies but also to the use of social networks and computer applications.

### Sample and population

Computer science students at the University of the Balearic Islands formed the population of the study, which is a four-year degree with 517 enrolled students in 2012/13, generating 3311 subject registrations.

Due to the necessity of using the same learning processes for all of the students in a group or subject, we could not do a random selection of students. Therefore, we decided to do a conglomerate study by selecting entire groups of students of a subject.

We selected 15 subjects with a total number of 482 students. The total number of subjects in the degree is 57. Our sample covered 26.32% of the subjects and 14.56% of the total registration. Considering the number of total student registrations (3311) and the number included in the study (482), the error of our test was $e = 0.042$.

### Experimental design

The experimental design determined that the selected groups of students were subjected to different strategies or types of motivation. In all cases, the motivation was focused on completing self-assessment tasks for exercises related to the subject. The number of students that completed the exercises and the percentage of them that did the self-assessment were gathered.

The independent variable or determining factor is a nominal variable with four different levels. These levels corresponded to each of the types of motivation used by the teacher over the student groups: initial motivation, regular motivation, rewarded motivation, and social motivation. In all cases, the students had one week to complete the self-assessment.

The first set of student groups corresponded to the initial motivation variable value. In this set, the students received only an initial explanation of how to perform the self-assessment and the benefits of this type of tasks. During the period open for the self-assessment, no other motivation or information about the process was given.

The second set was the regular motivation. The student groups in this set were continuously motivated to do the self-assessment. During the week assigned for the self-assessment, they usually received motivation from the teachers to do it and explanations about the importance and benefits of the self-assessment processes [11].

In the case of the rewarded motivation, the students that completed the self-assessment were rewarded with an increase in the mark of the correspondent activity. More concretely, the mark was increased by 10% for those students that completed the self-assessment task. Apart from that, they received only a first explanation of the self-assessment benefits, as in the case study of initial motivation.

Finally, the set of social motivation corresponded to the groups of students that were motivated to do the self-assessment through social and e-learning environments, such as social networks. The motivation was given only in the presentation of the activity, as in the case of the initial motivation, but all of the process took place on a social web site, more specifically, Facebook. All of the phases of the activities were published in a private group on Facebook. This process was also used for other activities and the contents of the subject during the course. The activities related to the contents of the subject were presented in a publication on the social network, and the results had to be uploaded to any cloud storage service and published in the Facebook group. Finally, the self-assessments were performed using a rubric in a form document on Google Drive, but the students had to publish the results of the rubric on Facebook. Thus, all of the students were able to know the achievements of their classmates.

The dependent variable or criterion was a quantitative variable with an indicator corresponding to the percentage of participation in the self-assessment task. The null hypothesis of the study was that all of the student groups show the same level of participation in the self-assessment tasks independent of the motivation received.

## Tools for the experiment

Our study on the student motivation was conducted using activities and tasks related to the self-assessment of the students. Therefore, a deeper analysis of the self-assessment tools and the social networks used in the experiment is necessary. Another important point was the homogeneity of the experiments. Different teachers did the experiments, so it was important to keep some homogeneity in the tools, processes and methods used in all of the groups.

One of the most popular tools for the process of self-assessment is the scoring rubric, or just called rubric. There are other alternatives for the self-assessment, such as tests, but rubrics are the most accepted ones. Rubrics relate the learning objectives with the level of accomplishment for each of these objectives [20]. To create a rubric, professors have to think about the main objectives that are part of the performance of an activity and to explicitly include them in the rubric list of objectives. They also have to clearly detail the different levels of accomplishment to allow the students to fix by themselves the level of achievement. The scoring rubric is a matrix in which the rows are the learning objectives and the columns the accomplishment level.

Other studies showed a web-based system in which a set of learning strategies were developed to motivate the students to perform the self and peer assessments and, in general, self-regulated learning [21]. From another point of view, self-assessment is a good way to motive students in their learning process. For example, Hattum Janssen et al. showed a correlation between the marks from the students and the teachers as well as an increase in the motivation of the students for first-year students [22].

Rubrics are the tools we decided to use for the self-assessment. Although all of the subjects and groups of students were in computer science, the contents of the subjects were quite heterogeneous, and consequently, the type of activities and exercises were very different, which was the main reason for the rejection of the use of the same rubric in all of the subjects. We finally decided to give some guidelines and support to the teachers involved in the study. Thus, they were able to create the most suitable rubrics for their particular exercises and activities. Then, an expert in the generation of rubrics checked and validated them.

The method of motivation for the students also had to be homogenized. As in the case of the creation of the rubric, some guidelines were given to the teachers to repeat similar motivation strategies. In the first case, the initial motivation, the professors took approximately 10 minutes to explain in the classroom with the students that self-assessment was important because they gained an appreciation of their weakness in the exercises. This explanation was given the same day that the teacher delivered the solutions for the problem/exercise. A paper copy of the rubric was also delivered. The teacher asked for the completed rubric to be given back after that week.

In the second type of motivation, regular one, the initial explanation of the benefits of the self-assessment was also given in the classroom the day of the solution delivery. This initial motivation was reinforced with regular motivations during the week of the study. These reinforcements were given during the lectures in the physical classroom. Each of them was about five minutes long, with a total of two reinforcements during the week.

The third strategy, the rewarded one, consisted of giving a bonus to the exercise mark to those students that performed the self-assessment. The initial motivation was also performed, but not the regular one. Instead of this reinforcement, the students obtained a 10% increase in the mark for the exercise for completing the rubric.

In the last type, the social one, the students used social web tools for all of the activities in the classes. The professors used a private Facebook group to publish the content of the course. The students were requested to enroll in the Facebook group, and they used it to publish a log of their activities, exercises, etc. The students were requested to publish a Facebook entry each time that they performed some task. The same happened when the teacher introduce new content, which was the way of working throughout the course. The self-assessment task was introduced during lecture time in the classroom, but the teacher also published an explanation of the benefits of the self-assessment in the Facebook group. The rubric was also published in the same method. There was no further reinforcement or motivation for the fulfilment of the self-assessment. The rubrics were delivered and collected using a form on Google Drive.

The objective was to study if the participation grade of the students using this last strategy is equal to some of the other ones. Percentages of participation for each strategy were compared, and statistical hypothesis testing was used to measure the significant differences between the types of motivation.

## 4. Results

All of the experiments were conducted using 15 student groups. The type of motivation was assigned randomly to each student group. Four student groups were included in the sets of initial, rewarded and social motivations and three in the regular set.

The data gathered from the student groups were the number of students that completed the activities and the number of those that did the self-assessment task. The results for the different groups are presented in Table 1. The results are presented individually for the student groups. The percentage of students that performed the self-assessment tasks was calculated in relation to the total number of activities delivered because it does not make sense to do an assessment of a non-delivered problem or activity. Each of those percentages was used as data for the study. The null hypothesis ($H_0$) was that the percentage of participation of the students was equal along the four experimental sets. The alternative hypothesis ($H_1$) of our study was that the number of students that performed self-assessment tasks was influenced by the type of motivation that they received from the teacher and, consequently, the percentage of participation showed significant differences among the experimental sets.

The dependent variable was ordinal, and the independent variable was nominal with four levels (motivation types). In these cases, the suitable statistical tests are the one-way analysis of variance (ANOVA) or the Kruskal-Wallis [23] test. The use of ANOVA is conditioned to having samples from a population that shows a normal distribution and experimental sets with similar variances. It is usual to consider that if the largest standard deviation exceeds twice the smallest, the ANOVA should be rejected. Table 4 shows that the standard deviation of the initial set was greater than twice the social one. Kruskal-Wallis is also used when the examined groups are of unequal size (different number of participants). ANOVA could not be used in our study. We chose the Kruskal-Wallis test for these reasons. The Kruskal-Wallis test indicates that there is a statistically significant difference among the sample sets if the value obtained in the test is higher than the value of the correspondent Chi-squared distribution ($\chi^2$) [24]. Table 2 shows samples from the four sets (types or strategies of motivations). These samples correspond to the students' participation percentages. The table also includes, in brackets, the ranges of the samples, i.e., their rank position considering all of the samples.

We performed the Kruskal-Wallis test using the R statistical software. We obtained $K=12.129$ as result of the test with a $\rho=0.007$. We needed to compare this value with the value of the Chi-squared distribution $\chi^2_{\alpha,g-1}$, where α is the desired significance or alpha level (we chose α=0.05) and g the number of sets (with g−1 the degrees of freedom of the $\chi^2$ distribution). We used a table of the chi-squared probability distribution 2 to obtain the value $\chi^2_{0.05,3} = 7.815$. As $K > \chi^2_{\alpha,g-1}$ the test was significant and a difference exists between at least two sets of the samples. It was necessary to repeat the test for each single pair of sets to detect the pair of sets with or without evidence of differences among the samples.

Table 3 shows the individual results of applying the Kruskal-Wallis test to a pair of sample sets, which is equivalent to applying the Mann-Whitney test. In these cases, the number of sets considered is 2, so we need to compare the results of the test ($K$) with the value of the chi-squared distribution $\chi^2_{0.05,1} = 3.841$. We can see that all of the values of the test are greater than 3.841, except for the Social-Regular pair. Therefore, there are significant differences in all cases, except for this last set.

Once the significant difference between the experimental sets is calculated, the results of the student participation should be calculated. Kruskal-Wallis is a non-parametric test and the normal distribution of the samples cannot be ensured. In these cases, the statistical test establishes the hypothesis in relation to the median instead of the mean. Therefore, it is better to analyze the results using medians and quartiles than using confidence intervals. Table 4 shows the statistics of the experimental sets: means, medians standard deviations and quartile boundaries. Box plots are the usual graphical representations for this type of data, but strip charts are better for our data set (Figure 1).

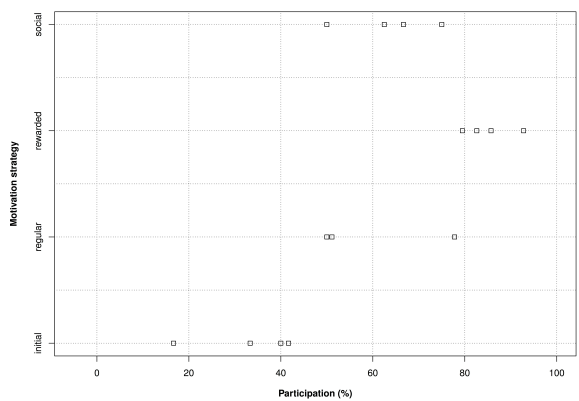

**Figure 1**. Students' participation strip chart for different motivation strategies

## 5. Discussion

In the first phase of our experiment, the statistical hypothesis test indicated that there were significant differences among the sets of samples. There is evidence that different types of motivation generate different percentages of participation by the students. This fact validates the idea that the student motivation to encourage them to self-assessment is very important.

The Kruskal-Wallis test is just able to ensure the significant difference at least between two experimental sets. Therefore, we need to test that difference between each pair of experimental sets. The results in Table 3 indicated a significant difference between the experiment sets, except for the case of regular and social motivation.

The real differences among the motivation strategies can be analyzed based on the results in Table 4. We used the Kruskal-Wallis test to ensure the differences among sets. This test establishes the validity of the hypothesis using the median instead of the mean. Therefore, it is more logical to compare the quartiles and the median among the experimental sets instead of the mean and the confidence intervals.

Table 1. Student's participation results

|  | Registered students | Activities completed | Self-assessments | Percentage (%) |
|---|---|---|---|---|
| Initial | 30 | 30 | 5 | 16.67 |
| Initial | 12 | 12 | 5 | 41.67 |
| Initial | 34 | 24 | 8 | 33.33 |
| Initial | 19 | 10 | 4 | 40.00 |
|  |  |  |  |  |
| Regular | 9 | 9 | 7 | 77.78 |
| Regular | 24 | 24 | 12 | 50.00 |
| Regular | 48 | 45 | 23 | 51.11 |
|  |  |  |  |  |
| Rewarded | 43 | 39 | 31 | 79.49 |
| Rewarded | 111 | 111 | 103 | 92.79 |
| Rewarded | 52 | 49 | 42 | 85.71 |
| Rewarded | 23 | 23 | 19 | 82.61 |
|  |  |  |  |  |
| Social | 28 | 24 | 16 | 66.67 |
| Social | 24 | 24 | 12 | 50.00 |
| Social | 16 | 12 | 9 | 75.00 |
| Social | 9 | 8 | 5 | 62.50 |
|  |  |  |  |  |

Table 2. Sets of samples for the Kruskal-Wallis test. Units: % (rank)

| Initial | Regular | Rewarded | Social |
|---|---|---|---|
| 16.67 (1) | 77.78 (11) | 79.49 (12) | 66.67 (9) |
| 41.17 (4) | 50.00 (5.5) | 92.79 (15) | 50.00 (5.5) |
| 33.33 (2) | 51.11 (7) | 85.71 (14) | 75.00 (10) |
| 40.00 (3) | - (-) | 82.61 (13) | 62.50 (8) |

Table 3. Results of the Kruskal-Wallis for pairs of experimental sets. Units: $K(\rho)$

|  | Initial | Regular | Rewarded | Social |
|---|---|---|---|---|
| **Initial** | | * | * | * | * |
| **Regular** | 4.500 (0.034) | | * | * | * |
| **Rewarded** | 5.333 (0.021) | 4.500 (0.034) | | * | * |
| **Social** | 5.333 (0.021) | 0.032 (0.858) | 5.333 (0.021) | | * |

Table 4. Statistics of participation for the motivation strategies. Units: %

|  | Initial | Regular | Rewarded | Social |
|---|---|---|---|---|
| Mean | 32.92 | 59.63 | 85.15 | 63.54 |
| Median ($P_{50}$) | 36.66 | 51.11 | 84.16 | 64.58 |
| Standard. desviation | 11.41 | 15.73 | 5.69 | 10.42 |
| $P_0$ | 16.67 | 50.00 | 82.61 | 50.00 |
| $P_{25}$ | 29.16 | 50.55 | 81.93 | 59.37 |
| $P_{75}$ | 40.42 | 64.44 | 87.48 | 68.75 |
| $P_{100}$ | 41.67 | 77.78 | 92.79 | 75.00 |

The strip chart representation is useful for comparing the dispersion of the samples of the data sets. Figure 1 shows that there were no overlaps between the values of the initial and rewarded sets with either of the other two motivation strategies (regular and social). In contrast, social and regular showed different medians and means, but their ranges are almost completely overlapped.

It is clear that higher student participation is obtained when they receive a reward from the process or the activity (median of 84.16%). Students are usually worried about their marks, and they are motivated to do some tasks when it can improve the final results in the form of a mark.

At the other end the lowest participation (median of 36.66%) was observed when the students received only the initial motivation without frequent reinforcement, reward or use of any technology. Once again, the importance in the motivation of the students is reflected in this partial result. This result was the experimental set with the lowest level of motivation from the teachers.

Finally, the regular and social experimental sets showed significant differences from the rest of the sets but not from each other. Despite this lack of difference, the medians and means of the student participation were slightly different: 51.11% and 64.58% in favor of the social set. In any case, the use of these two motivation types involved a higher participation of the students than in the case of only the initial motivation but a lower level than when the students received a mark increase.

It is quite logical that the case of regular motivation showed a higher participation than in the case of the initial one. The motivation is very important in the self-assessment process, and the students obtained a continuous one in this case. However, the results from the cases of the social sets are very interesting. In these cases, the students also obtained only the initial motivation. The differences were just the use of social networks and e-learning tools. Our results indicated that we can obtain the same result as a regular and continuous motivation by replacing it with the use of social networks and tools. Moreover, this result showed evidence that the social computing, social sites and e-learning environments are not only a tool by themselves but they are also a motivation for the students.

Social and regular motivations did not show significant differences. The same result will be obtained when they are used.

It is necessary to remark that the study was conducted on computer science students. These students are used to technological tools and the use of the latest advances in computers. However, the use of social networks and social computing is widespread, and we suggest that this result is not limited to students of computer science. In any case, the results are interesting in the field of computer science and in engineering studies.

In addition to the statistical discussion of the results, the opinions of the teachers that took part in the experiments were quite important. The general feeling of the teachers was in line with the statistical results. The professors think that the students participated more when they received an increase in marks, which is most likely a problem of the teachers in explaining the importance of self-assessment tasks. The students do some task when they know that it is an important for obtaining something. If they participate more when they obtain a mark increase, it is because it benefits them. If the students knew the importance of the self-assessment to their skills and learning processes, they would most likely have completed these tasks at a higher percentage.

The participation when social applications are involved in the learning process was very similar to the regular motivation. The feeling of the teachers was that the students were more motivated to perform tasks when they can share the experience in a social network. They encourage themselves when they can share and obtain feedback from other students. They had the feeling of a social and collective process of learning. After the experiments, some students were interviewed. They explained that when they used social applications they had extra motivation.

## 6. Conclusions

This study examined the effects of several motivation strategies on student participation. The study was focused on the completion of self-assessment tasks. Three traditional strategies (initial, regular and rewarded) were compared with the use of social tools and social networks. The group for the initial motivation was the control group and the increase in the percentage of student participation was compared to this group. The motivation with the highest participation was the rewarded one. The social and regular motivation strategies seemed to be very similar. In fact, the study did not show significant differences among them.

Despite this lack of difference, the median values showed a slight difference in favor of the social motivation

The results of the study indicated that different motivations resulted in different levels of participation in the self-assessment tasks. Moreover, the results showed that the students were motivated to do it using the social tools and social networks without receiving any other type of motivation or feedback. Thus, social networks are not just tools but also a way of motivating the students.

In future work, it would be interesting to determine if these results remain constant with students from other disciplines. Although the students of technical degrees are more used to new technological tools, the general feeling is that the use of social computing is widespread enough that other students will not have reservations to use this type of technology.

## Acknowledgments

It is our pleasure to express our thanks to Professor Joe Miró Julià because of his important help in the development of the experiment, his suggestions and his contributions that improved this manuscript.

This work was partially financed by the University of the Balearic Islands through the innovation and teaching improvement projects 2010-11, 2011-12 and 2012-13 (*Estudio de autoevaluacion y coevaluacion en el EEES*; *Estudio de motivacion y participacion del alumnado en actividades de autoevaluacion en los estudis del EEES*; and *Aplicaciones moviles para la mejora de la participacion de los alumnos en actividades en el aula*) and Departament de Ciències Matemàtiques i Informàtica UIB.